\documentclass[%
reprint,
 amsmath,amssymb,
 aps,
]{revtex4-2}

\usepackage{graphicx}
\usepackage{dcolumn}
\usepackage{bm}

\usepackage{subfigure}

\begin{document}

\preprint{APS/123-QED}

\title{Towards a Consistent Calculation of the Lunar Response to Gravitational Waves}

\author{Han Yan$^{1,2}$}
\author{Xian Chen$^{1,2}$}
  \email{Corresponding author. \\ xian.chen@pku.edu.cn}
\author{Jinhai Zhang$^{3}$} 
\author{Fan Zhang$^{4,5}$}
\author{Mengyao Wang$^{5}$} 
\author{Lijing Shao$^{2}$}

\affiliation{$^{1}$Department of Astronomy, School of Physics, Peking University, 100871 Beijing, China}

\affiliation{$^{2}$Kavli Institute for Astronomy and Astrophysics at Peking University, 100871 Beijing, China}

\affiliation{$^{3}$Institute of Geology and Geophysics, Chinese Academy of Sciences, Beijing 100029, China}

\affiliation{$^{4}$Institute for Frontiers in Astronomy and Astrophysics, Beijing Normal University, Beijing 102206, China}

\affiliation{$^{5}$Department of Astronomy, Beijing Normal University, Beijing 100875, China}

\date{\today}

\begin{abstract}
The recent increasing interest in detecting gravitational waves (GWs) by lunar
seismic measurement urges us to have a clear understanding of the response of
the moon to passing GWs.  In this paper, we clarify the relationship
between two seemly different response functions which have been derived
previously using two different methods, one taking the field-theory approach
and the other using the tidal force induced by GWs.  We revisit their
derivation and prove, by both analytical arguments and numerical calculations,
that the two response functions are equivalent. Their apparent difference can
be attributed to the choice of different coordinates.  Using the correct
response function,  we calculate the sensitivities (to GWs) of several designed
lunar seismometers, and find that the sensitivity curves between $10^{-3}$ and
$0.1$ Hz are much flatter than the previous calculations based on normal-mode model. Our results will help
clarifying the scientific objectives of lunar GW observation, as well as provide
important constraints on the design of lunar GW detectors.
\end{abstract}

\maketitle


\section{\label{sec:intro}Introduction }

The detection of gravitational waves (GWs) in the frequency window of $10-10^2$
Hz \cite{2016PhRvL.116f1102A} as well as nano-Hz
\cite{2023RAA....23g5024X,2023A&A...678A..50E,2023ApJ...951L...9A,2023ApJ...951L...6R}
encourages the efforts to detect GWs in other frequency bands.  Several
projects plan on using interferometers to detect milli-Hertz (mHz) GWs,
including the Laser Interferometer Space Antenna (LISA
\cite{2017arXiv170200786A}), TianQin \cite{2016CQGra..33c5010L}, and Taiji
\cite{2021PTEP.2021eA108L}.  There are also discussions and designs of interferometers to probe
deci-Hertz ($0.1$ Hz, or deci-Hz) GWs \cite{2020CQGra..37u5011A}, such as the
DECIGO \cite{2011CQGra..28i4011K} and TVLBAI \cite{2023arXiv231008183A}. 

An alternative approach is to take advantage of the quietness of the moon and
use it as a resonant GW detector \cite{1968PhT....21d..34W}.  However,
detecting the response of the moon to passing GWs requires a redesign of lunar
seismometers, so as to achieve particularly high sensitivities. The recent
studies based on several designs suggest that the sensitivity to GWs is the
best around deci-Hz, which will allow us to detect merging white dwarf binaries, intermediate-mass black hole binaries (IMBHBs) and super-massive black hole binaries (SMBHBs) out to cosmological
distances, as well as the GW background produced in the early universe
\cite{2021ApJ...910....1H,2023SCPMA..6609513L}.

One essential element in studying the lunar response to GW
is the calculation of the force density imposed by GW on an elastic body.
The theory was first laid down by Dyson (\cite{1969ApJ...156..529D}, hereafter Dy69). 
He started from the field theory and, by introducing a coupling term between GW and the elastic body, 
derived an external-force density 
\begin{eqnarray}
    \vec{f} =  -\nabla  \cdot \left ( \mu \mathbf{h} \right ) ~, \nonumber
\end{eqnarray}
where $\mu$ is the shear modulus and $\mathbf{h}$ refers to the 3-dimensional
spatial components of the GW tensor. Using this equation, Dyson studied the
response of an infinite half-space to a train of passing GW. Later, Ben-Menahem
applied the same force density to a more realistic lunar model, a radially
heterogeneous elastic sphere, and derived an analytical response solution
(\cite{1983NCimC...6...49B}, hereafter BM83).  A modern version of the
derivation can be found in Ref.~\cite{2019PhRvD.100d4048M} (hereafter Ma19).
These formulae derived in Dy69 and Ma19 form the basis for many later
calculations of the lunar or earth response to GWs
\cite{2014PhRvD..90d2005C,2014PhRvD..90j2001C,2014PhRvL.112j1102C,2023arXiv231211665K}.
They have also heavily influenced the scientific objectives of more recent
lunar GW projects, including the Lunar GW Antenna
(LGWA, \cite{2021ApJ...910....1H,2023SSRv..219...67B}).

An ambiguity, however, appears when one takes another viewpoint to calculate the 
lunar response function.
In the early studies of ground-based bar detectors
\cite{1995PhRvD..51.2517Z,1995PhRvD..52..591L,1996PhLA..213...16C,1996PhRvD..54.2409H,1996CQGra..13.2865B},
it is common to write the force density due to GW in the form of
\begin{eqnarray}
    \vec{f} = \frac{1}{2} \rho \frac{\mathrm{d}^{2} \mathbf{h} }{\mathrm{d} t^{2} }\cdot \vec{r}   ~, \nonumber
\end{eqnarray}
where $\rho$ is the mass density and $\vec{r}$ is the position vector. This
formula is usually called the ``tidal acceleration formula'', and has been
adopted by many textbooks (e.g., \cite{1973grav.book.....M,2009GReGr..41.1667H}). It is also used
in the recent science studies of lunar seismometer projects
\cite{2023SCPMA..6609513L}.  The apparent difference of this equation with
respect to the previous one naturally raises the question about which force
density should be used in the calculation of the lunar response.

This ambiguity has been noticed in several earlier papers.  In his attempt
to develop a fully general-relativistic treatment of the lunar response
\cite{1976SvPhJ..19..741D}, Dozmorov noticed the difference between the two
force densities \cite{1976SvPhJ..19..883D}. He attributed the difference to two
kinds of shear waves, one propagating at the speed of light and the other at
the speed of seismic wave. However, he did not give further explanation to the
cause or the relationship between these two waves. A recent review article
\cite{2019LRR....22....6H} (hereafter Ha19) also discusses the physical
meanings of the Dyson force and the tidal force by comparing the
surface displacement computed with Newtonian mechanics and the
displacement measured by an inertial sensor.  It argues that the two
displacements are equivalent, which hints that the two displacements are
connected by a change of the coordinate system, but a quantitative proof is
still lacking.  Probably because of the inconclusiveness of the previous
discussions, later works sometimes considered both types of forces and
presented two response functions. 

Inspired by these previous discussions, we decide to revisit the relevant
theories and try to resolve the apparent inconsistency caused by the
aforementioned two kinds of force densities. The paper is organized as follows.
In Section~\ref{sec:theo}, we review the dynamical equations of an elastic body
which is subject to the Dyson force or the tidal force, where we focus on
clarifying the physical difference between the coordinate systems in which the
equations are derived.  We then derive two response functions, corresponding to
the above two kinds of forces, and show that they can be unified into one
analytical formula.  In Section~\ref{sec:dis}, we apply our response functions
to a homogeneous isotropic sphere and a real lunar model, to show that in both
cases the numerical results agree with the analytical relationship.  We also
discuss the observability of GWs based on the appropriate response function.
Finally, in Section~\ref{sec:con}, we summarize our results and discuss their
implications for future lunar GW observation.  Throughout the paper  we will
adopt the International System of Units and the Minkowski metric $\eta _{\mu
\nu } = diag \left ( -1,1,1,1 \right ) $, unless otherwise mentioned.  Latin
alphabets represent three spatial indices, and Greek alphabets represent spacetime indices.

\section{\label{sec:theo}Theory}

This section reviews the dynamical equation of an elastic system in a GW field
with a flat spacetime background. Two sets of equations have been derived in
the literature due to different viewpoints, one based on the
transverse-traceless (TT) coordinate and the other in the lab frame. We will
first review the derivation of the equations, then clarify their mathematical
relation in the example of a radially heterogeneous elastic sphere.

\subsection{Equations in the TT coordinate}

When dealing with free particles moving in a GW field,
it is normally convenient to use the TT coordinate. 
In this coordinate system, which we denote by the subscript $A$, the line element reads
\begin{eqnarray}
    \mathrm{d}s^{2} = \left ( \eta _{\mu \nu } + h_{\mu \nu }^{TT}   \right )  \mathrm{d}x_{A}^{\mu } \mathrm{d}x_{A}^{\nu } ~.
\end{eqnarray}
Hereafter, for simplicity, we will omit the superscript $TT$ of the GW tensor $h_{\mu \nu }$.
The corresponding geodesic equation of a free, slowly moving particle is
\begin{eqnarray}
    \frac{\mathrm{d}^{2}  x_{A}^{i } }{\mathrm{d} t^{2} } = 0,
\end{eqnarray}
where $i=1,\,2,\,3$ denotes the three spatial directions.
In this equation, $h_{\mu \nu }$ does not appear, verifying the convenience of
using the TT coordinate.  
If, in addition, an electromagnetic (EM) force $f_{EM,A}^{i}$ is imposed on this particle, 
the equations of motion becomes
%
\begin{eqnarray}
    m\frac{\mathrm{d}^{2}  x_{A}^{i } }{\mathrm{d} t^{2} } = f_{EM,A}^{i},\label{eq:TTdy1}
\end{eqnarray}
where $m$ is the mass of the particle. Note that $f_{EM,A}^{i}$ should be expressed by the quantities in the TT coordinate.

Unlike a test particle, an elastic body consists of different parts, and their positions are
better described by a displacement field, $\vec{\xi}(t,\vec{x})$, which 
quantifies  the displacement of each part from the equilibrium position, $\vec{x}$.
Without GWs and external 
forces, the evolution of the body is governed by
\begin{eqnarray}
    \rho \frac{\partial^{2}  \xi^{i}  }{\partial t^{2} } = \frac{\partial \sigma ^{ij} }{\partial x^{j} }  ,\label{eq:elas1}
\end{eqnarray}
where $\sigma ^{ij}$ is the stress tensor for a locally homogeneous and isotropic medium. 
The tress tensor can be calculated with
\begin{eqnarray}
    \sigma ^{ij} = \lambda \delta ^{ij} \frac{\partial \xi ^{k} }{\partial x^{k} }  + \mu \left (\frac{\partial \xi ^{i} }{\partial x^{j} } + \frac{\partial \xi ^{j} }{\partial x^{i} }\right )  ,\label{eq:elas2}
\end{eqnarray}
where $\lambda$ and $\mu$ are two $\text{Lam}\acute{e}$ constants, and $\mu$ is also called  
the ``shear modulus''.

When GW is taken into account, Equation~(\ref{eq:elas1}) needs to be revised in
two aspects (noted on page 10 of Ha19). First, the meanings of $\vec{\xi}$ and $\vec{x}$ depend on the choice
of coordinate system. In particular, when the TT coordinate is considered,
$\vec{\xi}$ and $\vec{x}$ {\it do not} directly give the proper distance or
proper length, but differ from them by a small quantity of the order of $
\mathcal{O}(h)$.  Second, even when all the displacement vanishes, i.e.,
$\vec{\xi}(t,\vec{x})=0$ in TT coordinate, a shear force can still be induced
by the presence of GWs, because GWs change the proper distance between
different parts of the elastic body.

To account for these effects, a term should be added on the right-hand
side (RHS) of Equation~(\ref{eq:elas1}). 
As a result, Equation~(\ref{eq:elas1}), expressed in 
TT coordinate, becomes
\begin{eqnarray}
    \rho \frac{\partial^{2}  \xi_{A}^{i}  }{\partial t^{2} } = \frac{\partial }{\partial x_{A}^{j} } \left ( \sigma_{A} ^{ij} - \mu h^{ij}  \right ). \label{eq:elasGW}
\end{eqnarray}
This equation first appeared in Dy69, which started from a field-theory
approach, by writing down the interaction Lagrangian between GW and elastic
body. Compared to Equation~(\ref{eq:elas1}), the additional $\mu h^{ij}$ term
comes from the shear force induced by GWs. Equation~(\ref{eq:elasGW}) also
looks similar to Equation~(\ref{eq:TTdy1}). In fact, the RHS of Equation~(\ref{eq:elasGW})
calculates precisely the total EM force $\vec{f}_{EM,A}$ per unit volume that 
is driving the elastic body away from the geodesic.

In the case of lunar GW detection, the GW wavelength is usually much longer
than the size of the moon.  Therefore, the gradient of $h^{ij}$
over the entire body of the moon is small.  We can approximate the last
equation with
\begin{equation}
    \rho \frac{\partial^{2}  \xi_{A}^{i}  }{\partial t^{2} }  = \frac{\partial \sigma_{A} ^{ij}}{\partial x_{A}^{j} } -  \frac{\partial \mu }{\partial x_{A}^{j} } h^{ij}. \tag{6$'$}\label{eq:elasGW1}
\end{equation}

\subsection{Equations in the lab coordinate}

Another coordinate system which is commonly used in textbooks and papers to
describe the influence of GW is the ``lab coordinate''. It is also known as
the ``proper coordinate'' because the spatial components are defined using the
proper distance.  The line element in this coordinate is 
\begin{eqnarray}
    \mathrm{d}s^{2}  = && \eta _{\mu \nu }  \mathrm{d}x_{B}^{\mu } \mathrm{d}x_{B}^{\nu } \nonumber\\ && + \mathcal{O} \left ( 1 \right )\times  \left ( R_{\mu l \nu m}x_{B}^{l} x_{B}^{m}  \right )  \mathrm{d}x_{B}^{\mu } \mathrm{d}x_{B}^{\nu } + ...  ~,
\end{eqnarray}
where $R_{\mu l \nu m}$ is the Riemann curvature tensor, and we have used the
subscript $B$ to denote this coordinate.  The factor of $\mathcal{O}(1)$ can be found in
textbooks (e.g. Ref.~\cite{1973grav.book.....M}).  Meanwhile, we have omitted
the terms due to inertial acceleration and rotation of the lab frame 
because these effects due to the orbital motion and rotation of the moon 
appear at much lower frequencies than mHz.
Given this simplification, the leading terms of the metrics in the coordinate 
systems $A$ and $B$ are exactly
the same. In this way, the coordinates of the equilibrium positions have the same
values, no matter which coordinate we choose.

In the lab coordinate $B$, the geodesic equation of a free, slow-moving particle can be written as
\begin{eqnarray}
    \frac{\mathrm{d}^{2}  x_{B}^{i } }{\mathrm{d} t^{2} } = \frac{1}{2} \frac{\mathrm{d}^{2}  h^{i}_{j} }{\mathrm{d} t^{2} } x_{B}^{j} . 
\end{eqnarray}
The RHS is normally interpreted as a tidal force induced by GWs.  Note that the
components of the GW tensor in this equation can be made to appear identical to that in
Equation~(\ref{eq:elasGW1}) in the linear order, because there are sufficient residual freedom in two gauge choices \cite{1973grav.book.....M}, even though these two formulae are derived in different coordinate bases. 
With an additional EM force, $\vec{f}_{EM,B}$, the equation of motion becomes
\begin{eqnarray}
    m\frac{\mathrm{d}^{2}  x_{B}^{i } }{\mathrm{d} t^{2} } = f_{EM,B}^{i} +\frac{1}{2} m\frac{\mathrm{d}^{2}  h^{i}_{j} }{\mathrm{d} t^{2} } x_{B}^{j}.    \label{eq:GWtid}
\end{eqnarray}
One can compare it with Equation~(\ref{eq:TTdy1}) to see the consequence of choosing different
coordinates. 

To use Equation~(\ref{eq:GWtid}) on elastic bodies, we notice that the first
term on the RHS can be readily replaced by Equations~(\ref{eq:elas1}) and
(\ref{eq:elas2}), because they are already constructed using the proper distance.
The second term does not depend on the property of an elastic body, and hence
remains in the equation. Therefore, we derive 
\begin{eqnarray}
    \rho \frac{\partial^{2}  \xi_{B}^{i}  }{\partial t^{2} } = \frac{\partial  \sigma_{B} ^{ij}}{\partial x_{B}^{j} } +  \frac{1}{2} \rho\frac{\mathrm{d}^{2}  h^{i}_{j} }{\mathrm{d} t^{2} } x_{B}^{j}. \label{eq:elasGW2}
\end{eqnarray}

Equations~(\ref{eq:elasGW1}) and (\ref{eq:elasGW2}) clearly show the difference
caused by different coordinates. However, both equations describe the exact
same dynamics. To see this equivalence, it is important to understand that the
equilibrium positions, defined as zero-displacement position $\xi_{A/B} = 0$, are physically different, even though numerically they may appear
the same.  More specifically, in coordinate $B$ (lab frame) the equilibrium
position maintains the same {\it proper} distance from the origin of the
coordinate system. However, in coordinate $A$ (TT frame), the equilibrium
position changes its proper distance from the origin, while it is the
coordinate distance (i.e., $\Delta x^{i}_{A}$) that is kept constant.  

Because of this difference, the equilibrium point
in coordinate $B$ is accelerating with respect to the equilibrium point 
in $A$. The acceleration is $(d^2h_j^i/dt^2)x^j_{\rm eq}/2$ when measured in
coordinate $B$,
where $x_{\rm eq}^{j}$ is the coordinate of the equilibrium point. 
Notice that the values of 
$x_{\rm eq}^{j}$ in coordinate $A$ and $B$ differ by a small term of the order of
$\mathcal{O}(h)$.
Since the displacement fields $\vec{\xi} _{A} $
and $\vec{\xi} _{B}$ are defined relative to their respective equilibrium
points, and by the rule of addition of acceleration, we have  
\begin{eqnarray}
	 \frac{\partial^{2}  \xi_{B}^{i}}{\partial t^{2} }= \frac{\partial^{2}  \xi_{A}^{i}}{\partial t^{2} } +  \frac{1}{2} \frac{\mathrm{d} ^{2}  h^{i}_{j}}{\mathrm{d} t^{2} }   x_{\rm eq}^{j}+\mathcal{O}(h^2). \label{eq:relat}
\end{eqnarray}
The last term $\mathcal{O}(h^2)$ 
comes from a higher-order correction of the tidal force.
 
Finally, by combining Equations~(\ref{eq:elasGW1}), (\ref{eq:elasGW2}) and (\ref{eq:relat}), we find that
\begin{eqnarray}
    \frac{\partial \sigma_{A} ^{ij}}{\partial x_{A}^{j} } -  \frac{\partial \mu }{\partial x_{A}^{j} } h^{ij}  
&=&\frac{\partial  \sigma_{B} ^{ij}}{\partial x_{B}^{j} } +  \frac{\rho}{2} \frac{\mathrm{d}^{2}  h^{i}_{j} }{\mathrm{d} t^{2} } \left ( x_{B}^{j} - x_{\rm eq}^{j}  \right ) \nonumber \\
	&=& \frac{\partial  \sigma_{B} ^{ij}}{\partial x_{B}^{j} } + \mathcal{O}(h^{2}).\label{eq:fAB}
\end{eqnarray}
This equation can be considered as a projection of the same EM force onto two
different sets of base vectors.  In the following, we will omit the last term
of the order of $\mathcal{O}(h^{2})$.

\subsection{Analytical solutions for a radially heterogeneous elastic sphere}

A radially heterogeneous elastic sphere is a good first-order approximation of the real
structure of the moon.  Its surface response to GWs was first calculated
analytically in BM83 using Equation~(\ref{eq:elasGW1}), and later in Ma19 in
more detail. Here we mainly follow the convention in Ma19 unless mentioned
otherwise.

In Ma19, GW is described by a plane wave
\begin{equation}
	\mathbf{h} =  \Re \left \{ h_{0} \epsilon _{ij} e^{i\left ( \omega _{g}t - \vec{k}_{g}\cdot \vec{r}     \right ) }    \right \},\label{eq:GW}
\end{equation}
where $h_0$ is the GW amplitude, $\omega _{g}$ is the GW angular frequency,
$\vec{k}_{g} =\left ( 0,0,\omega _{g}/c\right )$ is the wave vector which we assume pointing in the
$z$ direction, and
\begin{eqnarray}
\epsilon _{ij} && = \begin{bmatrix}
 1 & 1 & 0\\
 1 & -1 & 0\\
 0 & 0 & 0
	\end{bmatrix}
\end{eqnarray}
is the polarization tensor. This is identical to the $e= \lambda = \nu =0$ case in Ma19.  Notice that it is a special case in which the GW is linearly polarized. If it is in other polarization states,
an $\mathcal{O}(1)$ modification should be made to the polarization-dependent part of the result (i.e., $f^{m}$ in Equation~\ref{eq:surfsol}). We then neglect the
$\vec{k}_{g}\cdot \vec{r}$ term in Equation~(\ref{eq:GW}) because $\omega
_{g}R/c \ll 1$, where $R$ is the radius of the sphere.

Given the above GW, the analytical solution to Equation~(\ref{eq:elasGW1}) is
\begin{eqnarray}
    \vec{\xi }^{A} _{k} \left ( \theta ,\varphi ,t \right ) = h_{0} \vec{s } _{k} \left ( \theta ,\varphi  \right ) \Re \left \{ \bar{g}_{n}\left ( t \right ) \right \}   f^{m} \alpha^{A} _{2n} ~,  \label{eq:surfsol}
\end{eqnarray}
where $k=nlm$, $\left | m \right | \le l$, and for GWs only the spheroidal modes of $l=2$ 
are excited \cite{1983NCimC...6...49B,1996CQGra..13.2865B}. The term $\vec{s } _{k}$
is the displacement eigenfunction of the spherical modes:
\begin{eqnarray}
    \vec{s } _{k} \left ( \theta ,\varphi  \right ) = && U_{2n} \left ( R \right )  \mathcal{Y}_{2m} \left ( \theta ,\varphi  \right ) \hat{e}_{r} \nonumber \\ && + \frac{1}{\sqrt{6} } V_{2n} \left ( R \right ) \partial _{\theta }  \mathcal{Y}_{2m} \left ( \theta ,\varphi  \right ) \hat{e}_{\theta }  \nonumber \\
    && + \frac{1}{\sqrt{6} } V_{2n} \left ( R \right ) \left ( \sin \theta  \right ) ^{-1}  \partial _{\varphi  }  \mathcal{Y}_{2m} \left ( \theta ,\varphi  \right ) \hat{e}_{\varphi   }.
\end{eqnarray}
The spherical coordinates here, $\theta$ and $\varphi$, mark the position on the surface of the sphere. 
The function $\bar{g}_{n}$ is called the source-time function, and it can be calculated with
\begin{eqnarray}
    \bar{g}_{n}\left ( t \right ) =  \frac{e^{i\omega _{g}t} }{\omega _{n}^{2} - \omega _{g}^{2} + i\omega _{n}\omega _{g}/Q_{n} } . \label{eq:stf}
\end{eqnarray}

Note that our $\bar{g}_{n}$ is slightly different from those in BM83 and Ma19 because of our choice of normalization, and we have verified it with our numerical calculations. 
The dependence on the wave vector $\vec{k}_{g}$ and polarization of GW is contained in
\begin{eqnarray}
    f^{m} && = f^{m}\left ( e= \lambda = \nu =0 \right )  \nonumber \\
    && =  4\sqrt{\frac{\pi }{15} } \times \left ( \delta_{m, 2}+  \delta_{m,-2} \right ). 
\end{eqnarray}
Finally, $\alpha^{A}_{2n}$ depends on the radial structure of the sphere, 
\begin{eqnarray}
    \alpha^{A}_{2n} = &&  - \frac{\int^{R+}_{0} \frac{\partial \mu}{\partial r}\left(U_{2n}(r)+\frac{3}{\sqrt{6}} V_{2n}(r)\right) r^{2} d r}{\int^{R}_{0} \left(U^{2}_{2n}(r)+ V^{2}_{2n}(r)\right)\rho \left ( r \right )  r^{2} d r} \nonumber \\ = && \frac{\mu(R) R^{2}\left(U_{2n}(R)+\frac{3}{\sqrt{6}} V_{2n}(R)\right)}{\int^{R}_{0} \left(U^{2}_{2n}(r)+ V^{2}_{2n}(r)\right)\rho \left ( r \right )  r^{2} d r}  \nonumber \\ 
    && - \frac{\int^{R}_{0} \frac{\partial \mu}{\partial r}\left(U_{2n}(r)+\frac{3}{\sqrt{6}} V_{2n}(r)\right) r^{2} d r}{\int^{R}_{0} \left(U^{2}_{2n}(r)+ V^{2}_{2n}(r)\right)\rho \left ( r \right )  r^{2} d r}   ~.\label{eq:alp_Dy}
\end{eqnarray}
The upper integration limit $R+$ in the first line means that the integration
should be taken till the outer side of the surface. Notice that $\alpha_{2n}$
can be negative but it is real, as long as $U_{2n}(r)$ and $V_{2n}(r)$ are
real. The denominator in Equation~(\ref{eq:alp_Dy}) comes from the
normalization of $U^{2}_{2n}+ V^{2}_{2n}$ in Ma19.

To derive the solution to Equation~(\ref{eq:elasGW2}),
we find that the difference between Equations~(\ref{eq:elasGW1}) and (\ref{eq:elasGW2}) 
lies in the expression of the force density. Therefore, we can get the solution 
$\vec{\xi }^{B} _{k} \left ( \theta ,\varphi ,t \right )$
by replacing $\partial \mu / \partial r$ in Equation~(\ref{eq:alp_Dy})
with $\rho r\omega _{g}^{2}/2 $, the tidal-force density in the frequency domain.
Then we have
\begin{eqnarray}
     \alpha_{2n}^{B}  =   - \frac{\omega _{g}^{2}}{2} \frac{\int^{R}_{0}  \left(U_{2n}(r)+\frac{3}{\sqrt{6}} V_{2n}(r)\right)\rho \left ( r \right )  r^{3} d r}{\int^{R}_{0} \left(U^{2}_{2n}(r)+ V^{2}_{2n}(r)\right)\rho \left ( r \right )  r^{2} d r} ~.\label{eq:alp_tid}
\end{eqnarray}
Notice that this equation no longer contains the first term in
Equation~(\ref{eq:alp_Dy}), because the term comes from a discontinuity of
the gradient of shear modulus at the surface. When the tidal-force density is
concerned, an integration over the surface does not lead to such a term. 

\subsection{Relation between the two analytical solutions}

The solutions $\vec{\xi} _{k}^{A}$ and $\vec{\xi} _{k}^{B}$ derived above should,
according to Equation~(\ref{eq:relat}), satisfy the following relationship,
\begin{eqnarray}
	\vec{\xi} ^{A} = \vec{\xi} ^{B} -  \frac{1}{2} \mathbf{h}   \cdot \vec{R}  ~,\label{eq:sum}
\end{eqnarray}
where $\vec{R} = R\left ( \sin \theta \cos \varphi ,\sin \theta \sin \varphi ,\cos \theta  \right ) $, and $\vec{\xi } \equiv \sum_{n,m}\vec{\xi } _{k}$.

In what follows, it is more instructive to write $\vec{\xi }$ in terms of its three
spatial components,
\begin{eqnarray}
    \vec{\xi } \left ( \theta ,\varphi ,t \right )  = && h_{0} \cos \left ( \omega _{g}t  \right ) \times \nonumber\\ &&\Bigg[ T_{r} \sum_{m}  f^{m}   \mathcal{Y}_{2m} \left ( \theta ,\varphi  \right ) \hat{e}_{r} \nonumber \\ && + T_{h} \sum_{m}  f^{m} \partial _{\theta }  \mathcal{Y}_{2m} \left ( \theta ,\varphi  \right ) \hat{e}_{\theta }  \nonumber \\
    &&  +  T_{h} \sum_{m}  f^{m}  \frac{\partial _{\varphi  }  \mathcal{Y}_{2m} \left ( \theta ,\varphi  \right )}{\sin \theta } \hat{e}_{\varphi   } \Bigg] ~,
\end{eqnarray}
where we define the radial and horizontal response functions respectively as
\begin{equation}
    T_{r} \equiv \sum_{n} U_{2n} \alpha _{2n}\Re \left \{ \bar{g}_{n}\left ( t \right )e^{-i\omega _{g}t }  \right \}
\end{equation}
and
\begin{equation}
    T_{h } \equiv \sum_{n} \frac{V_{2n}}{\sqrt{6} }  \alpha _{2n}\Re \left \{ \bar{g}_{n}\left ( t \right )e^{-i\omega _{g}t }  \right \} ~.
\end{equation}

We also find that (see Appendix \ref{app:GWproj} for a proof)
\begin{eqnarray}
    \frac{1}{2} \hat{e}_{r} \cdot   \mathbf{h}   \cdot \vec{R} && = \frac{R}{2} h_{0}  \cos \left ( \omega _{g}t  \right )   \sum_{m} \mathcal{Y}_{2m} \left ( \theta ,\varphi  \right )f^{m}, \nonumber \\
    \frac{1}{2} \hat{e}_{\theta } \cdot   \mathbf{h}   \cdot \vec{R} &&= \frac{R}{4} h_{0}  \cos \left ( \omega _{g}t  \right )   \sum_{m} \partial_{\theta }  \mathcal{Y}_{2m} \left ( \theta ,\varphi  \right )f^{m}, \nonumber \\
    \frac{1}{2} \hat{e}_{\varphi } \cdot   \mathbf{h}   \cdot \vec{R} &&= \frac{R}{4} h_{0}  \cos \left ( \omega _{g}t  \right )   \sum_{m} \frac{\partial _{\varphi  }  \mathcal{Y}_{2m} \left ( \theta ,\varphi  \right )}{\sin \theta } f^{m}  ~. \label{eq:GWproj}
\end{eqnarray}

Using the above equations, we can
eliminate the common factors of Equation~(\ref{eq:sum}). Finally, we find
that
\begin{eqnarray}
    T_{r}^{A} &&= T_{r}^{B} - \frac{1}{2} R \nonumber \\
    T_{h}^{A} &&= T_{h}^{B} - \frac{1}{4} R ~. \label{eq:TFrela}
\end{eqnarray}
Notice that we have already used the following equation coming from the definition of $\bar{g}_{n}\left ( t \right )$, expect for a small resonant frequency region $\left | \omega _{g} - \omega _{n} \right | < \omega _{n}/Q_{n} $:
\begin{eqnarray}
    \Re \left \{ \bar{g}_{n}\left ( t \right )e^{-i\omega _{g}t }  \right \} \cos \left ( \omega _{g}t \right )  = \Re \left \{ \bar{g}_{n}\left ( t \right )  \right \}. 
\end{eqnarray}
We also clarify here that the different factors $1/2$ and $1/4$ in
Equation~(\ref{eq:TFrela}) come from the different factors on the RHS of
Equation~(\ref{eq:GWproj}), reflecting the behaviors of spherical harmonics. We will verify Equation~(\ref{eq:TFrela}) in
the next section by numerical calculations.

Several previous
works tend to take $T^{B}$ as the response function (e.g.,
\cite{2021ApJ...910....1H,2023SCPMA..6609513L})
and use it to infer the detectability of GWs. 
Here we would like to point out that $T^{B}$ is not directly proportional to the
readout of a seismometer, and hence should be used with caution.
The reason is given in the next section. 

\section{\label{sec:dis}Numerical test and application}

In this section,  we first calculate the response functions for a simplified
model, a homogeneous isotropic sphere, to verify our theory.  Then we employ a more realistic lunar model to derive more realistic response functions. Based on these
results,  we discuss the implication for lunar GW detection. In the calculation,
we use the MINEOS software package \cite{1988MINEOS} to calculate the spheroidal
eigenfunctions $U_{2n}(r)$ and $V_{2n}(r)$. We note that our $V_{2n}$ is a
factor of $\sqrt{6}$ smaller than that given by MINEOS because of a
different normalization, according to the annotation in MINEOS.

\subsection{Homogeneous isotropic sphere}

We first consider a homogeneous isotropic sphere, with the radius
$R=1000\mathrm{km}$, density $\rho = 3000\mathrm{kg\cdot m^{-3}}$, compressive wave speed $v_{p} = 8
\mathrm{km\cdot s^{-1}}$, shear wave speed $v_{s} = 4 \mathrm{km\cdot s^{-1}}$, and quality factor $Q=1000$. These values qualitatively reflect the ``averaged" properties of the moon. We choose two different
numbers of layers in MINEOS, $N_{L} = 200$ and $2000$. All the eigenfunctions for
the first $400$ normal modes (i.e., $0\le n \le 400$) are calculated.

\begin{figure}
\includegraphics[width=0.96\linewidth]{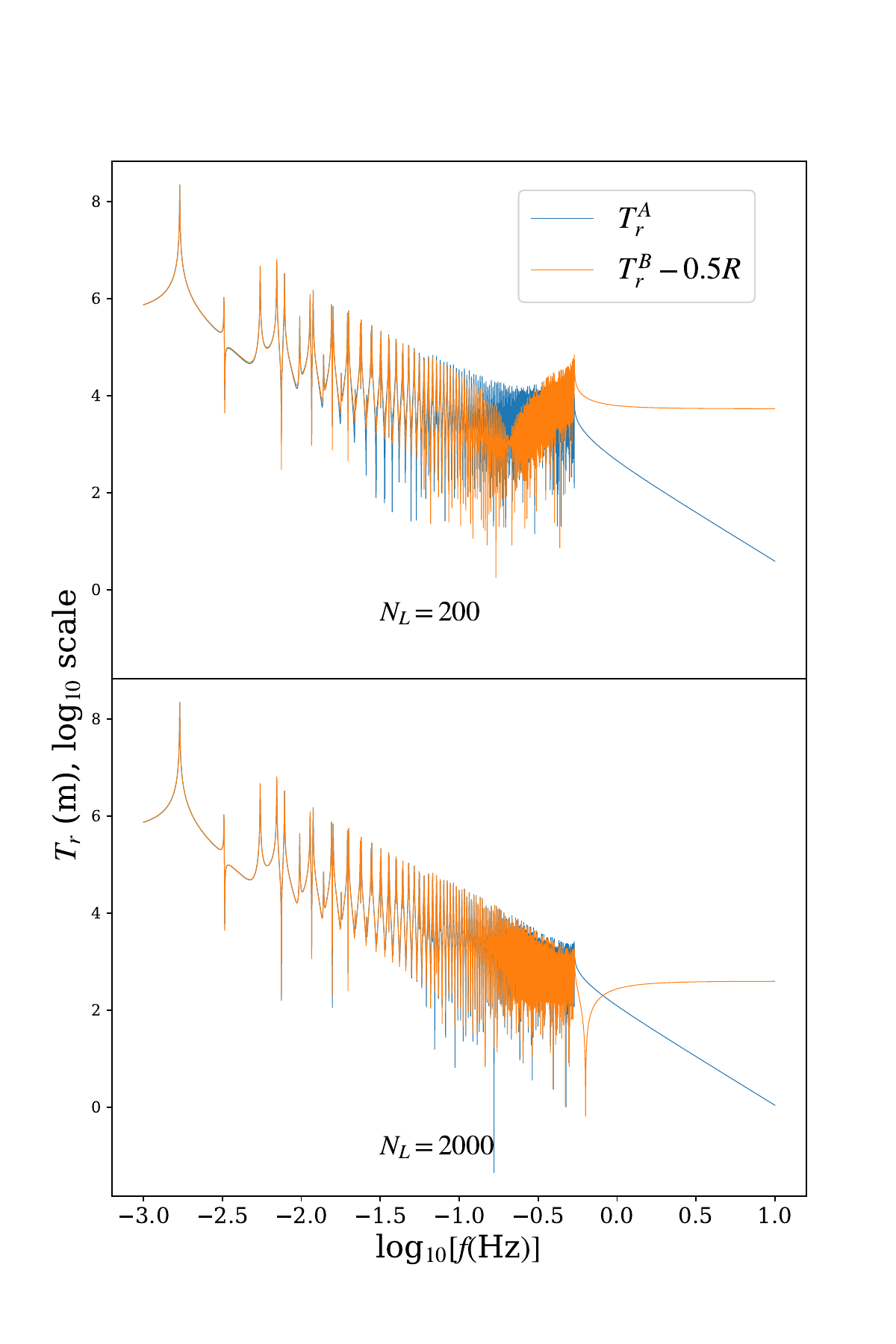}
\caption{\label{fig:1}Radial response functions $T_{r} (f)$ of a homogeneous isotropic sphere
calculated using different numbers of layers $N_{L}$. Notice that both axes are in log$_{10}$ scale.
The results corresponding to the Dyson force ($T_{r}^{A}$) and tidal force ($T_{r}^{B}-0.5R$) are shown in the same plot, for easier comparison.}
\end{figure}

The resulting radial response functions are shown in Figure~\ref{fig:1}.
The plot shows that Equation~(\ref{eq:TFrela}) works better at lower
frequencies, and also better when larger number of layers are used in the
calculation. The improvement with respect to the number of layers can be 
understood as follows.
When more layers are included, those eigenfunctions
$U_{2n}(r)$ and $V_{2n}(r)$ with larger $n$ can be more accurately
calculated.
Since larger $n$ corresponds to higher eigen-frequencies $\omega _{n} $,
the response function at higher frequency also becomes more accurate. The
mismatch at high frequency ($> 10^{-0.3}$ Hz) in the lower panel is mainly
caused by our truncation of normal modes at $n = 400$. We also get similar
results for the horizontal response function $T_{h}$, which is not shown here.

\subsection{\label{sec:compare}Real lunar model}

\begin{figure}
\includegraphics[width=0.95\linewidth]{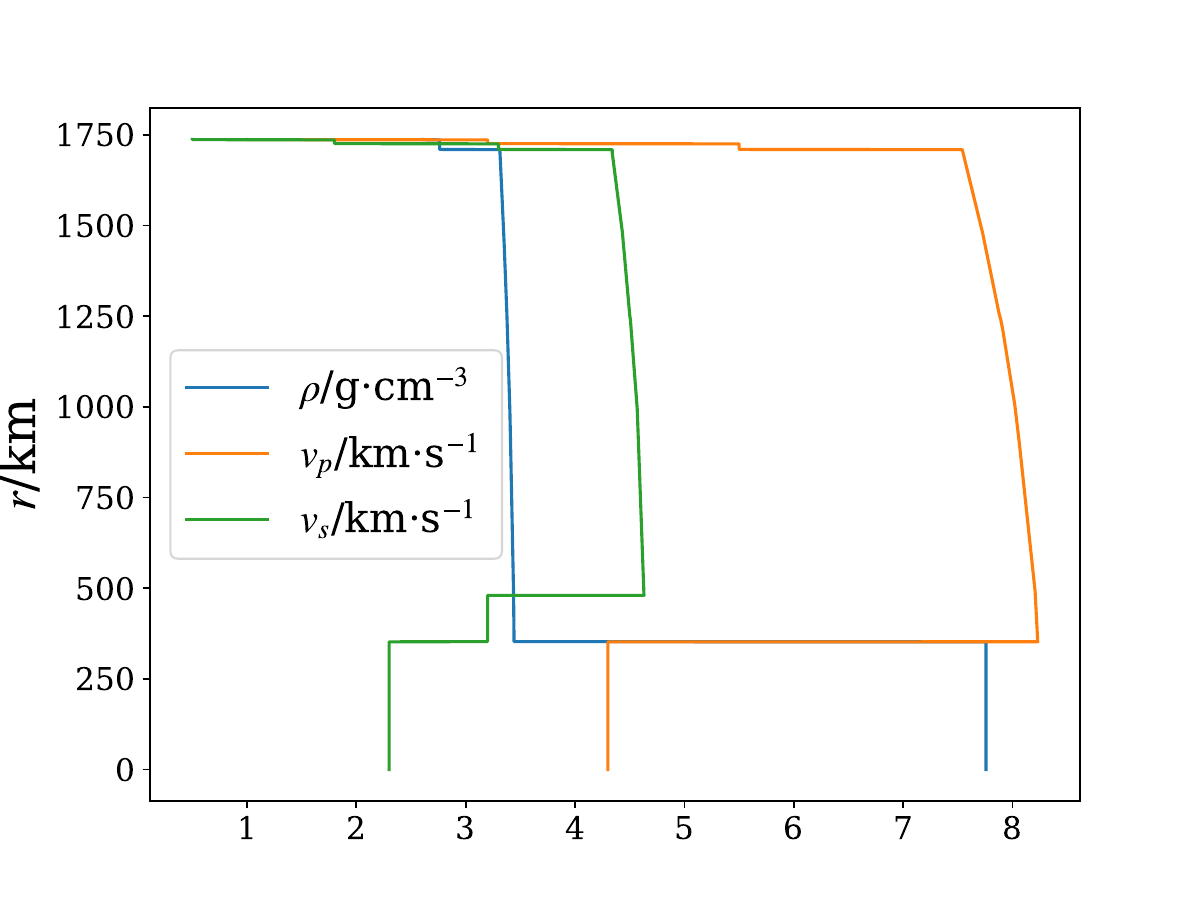}
\caption{\label{fig:2} 
Realistic lunar model used in this paper.  
The units are chosen so that three curves can be plotted using one $x-$axis.}
\end{figure}

\begin{figure}
\includegraphics[width=0.93\linewidth]{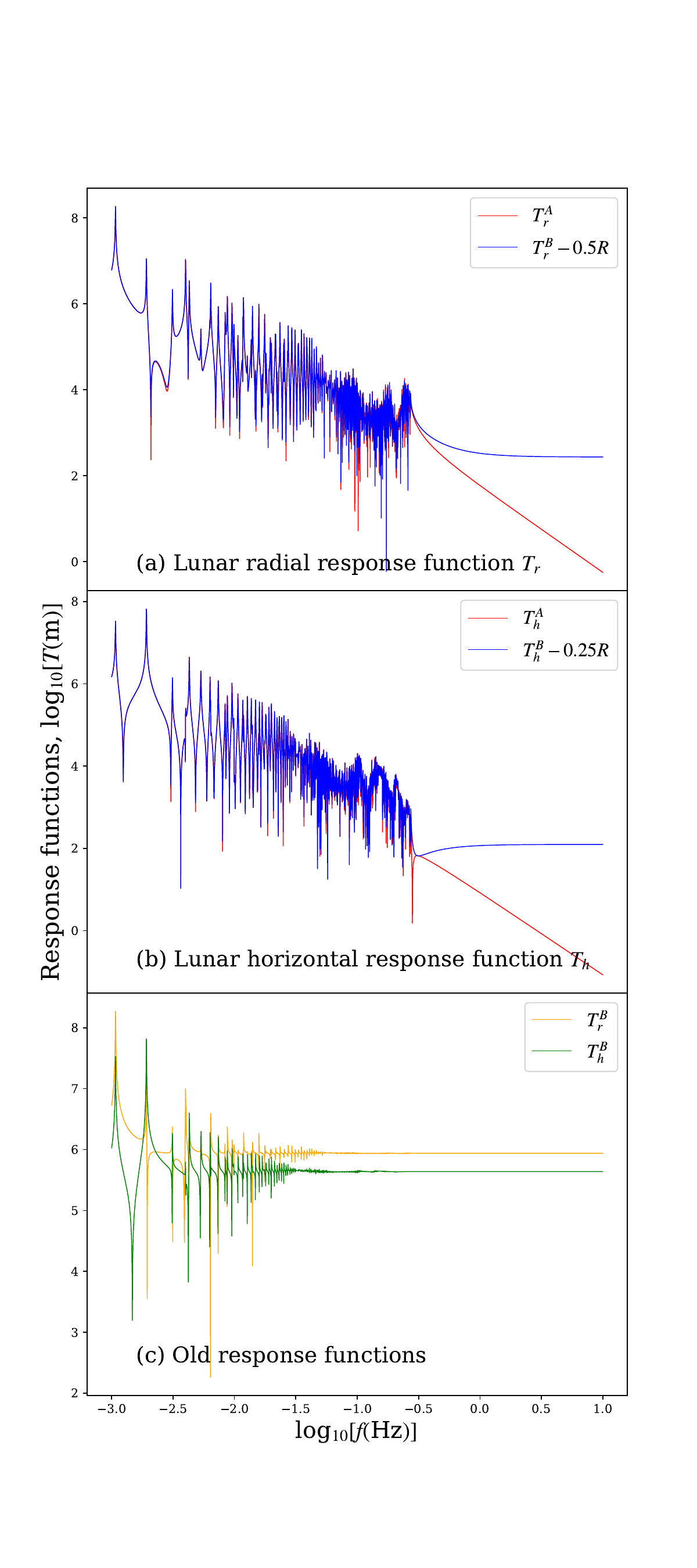}
\caption{\label{fig:3}Response functions for the realistic lunar model. 
(a) Radial response functions $T_{r} (f)$ derived from the Dyson force (A) and tidal force (B). 
(b) Horizontal response functions $T_{h} (f)$ for the two forces. 
(c) Old response functions $T_{r}^{B}$ and $T_{h}^{B}$, which are derived following the 
	method given in \cite{2021ApJ...910....1H}.}
\end{figure}

Figure~\ref{fig:2} shows the structure of our realistic lunar model.  It is
compiled from several published works
\cite{2011Sci...331..309W,2011PEPI..188...96G,2023Natur.617..743B}, so that we
can prepare a full input file for MINEOS. The model only has a
homogeneous core, and has a low velocity zone (LVZ) outside the core
\footnote{The parameters of the LVZ come from J. Zhang and his collaborators,
while the related paper is still in preparation.}. To increase the accuracy at
high frequencies, we generate 28501 layers by interpolating with the
original model data. The original and interpolated model files can be found in \cite{webs}.

The most uncertain part is the $Q$ value of the  lunar core. We
have run tests by changing $Q_{\rm core}$ between 200 and 5000. The results do not show
characteristic difference, so the exact value of $Q$ should not affect the main
conclusions of this work. We choose $Q_{\rm core} = 1000$ in our model. 

The radial and horizontal response functions derived for the interpolated model
are shown in Figure~\ref{fig:3}, panels (a) and (b). We have truncated the
normal modes at $n=400$. The value of the eigen-frequencies $\omega_{n}$,
quality factors $Q_{n}$ and response functions $T^{A}_{r/h}(f)$ can be found in \cite{webs}.  At the frequencies lower than $10^{-0.6} $
Hz, we see a good agreement between the results derived from the Dyson force
and the tidal force, which proves the validity of Equation~(\ref{eq:TFrela}). The
disagreement at higher frequencies, again, is caused by the artificial
truncation of normal modes in the calculation.

To compare with the response functions presented in earlier works, we also plot
in Figure~\ref{fig:3} (c) our $T_{r}^{B}$ and $T_{h}^{B}$, but without
subtracting, respectively, $0.5R$ and $0.25R$. The results recover those ``old"
response functions (e.g., presented in Figure~1 of
Ref.~\cite{2021ApJ...910....1H}). Comparing these old response functions with
those new ones in panels (a) and (b), we find the difference at high
frequencies, especially around $0.1 \text{Hz}$, which was previously considered
to be the sweet spot of a lunar GW detector.

\subsection{\label{sec:GW}Observables and detectability}

To evaluate the detectability of GWs by future lunar seismology projects, we
must first understand what is observable.  We emphasize here that it is the
acceleration $\partial^{2}  \xi_{A}^{i} / \partial t^{2} $, not $\partial^{2}
\xi_{B}^{i} / \partial t^{2} $, that is a direct observable for a seismometer.  
Generally
speaking, the former is measuring the {\it local} acceleration caused by EM forces, while the
latter also counts for the tidal acceleration with respect to the center of the moon. The local
acceleration is a quantity that an accelerometer, such as that installed in a
lunar seismometer or gravimeter, can directly measure. The latter {\it tidal}
acceleration, however, is not in simple proportion to the readout of
accelerometer. For example, consider two nearby test particles freely floating
in a vacuum. When GWs pass by, the proper length between the two particles will
vary, which induces a {\it tidal} acceleration. But each particle actually
follows its own geodesic motion (i.e., in free fall), so by the equivalence
principle any small-sized instrumentation, such as an accelerometer, attached
to either particle will give zero readout.

\begin{figure}
\includegraphics[width=0.95\linewidth]{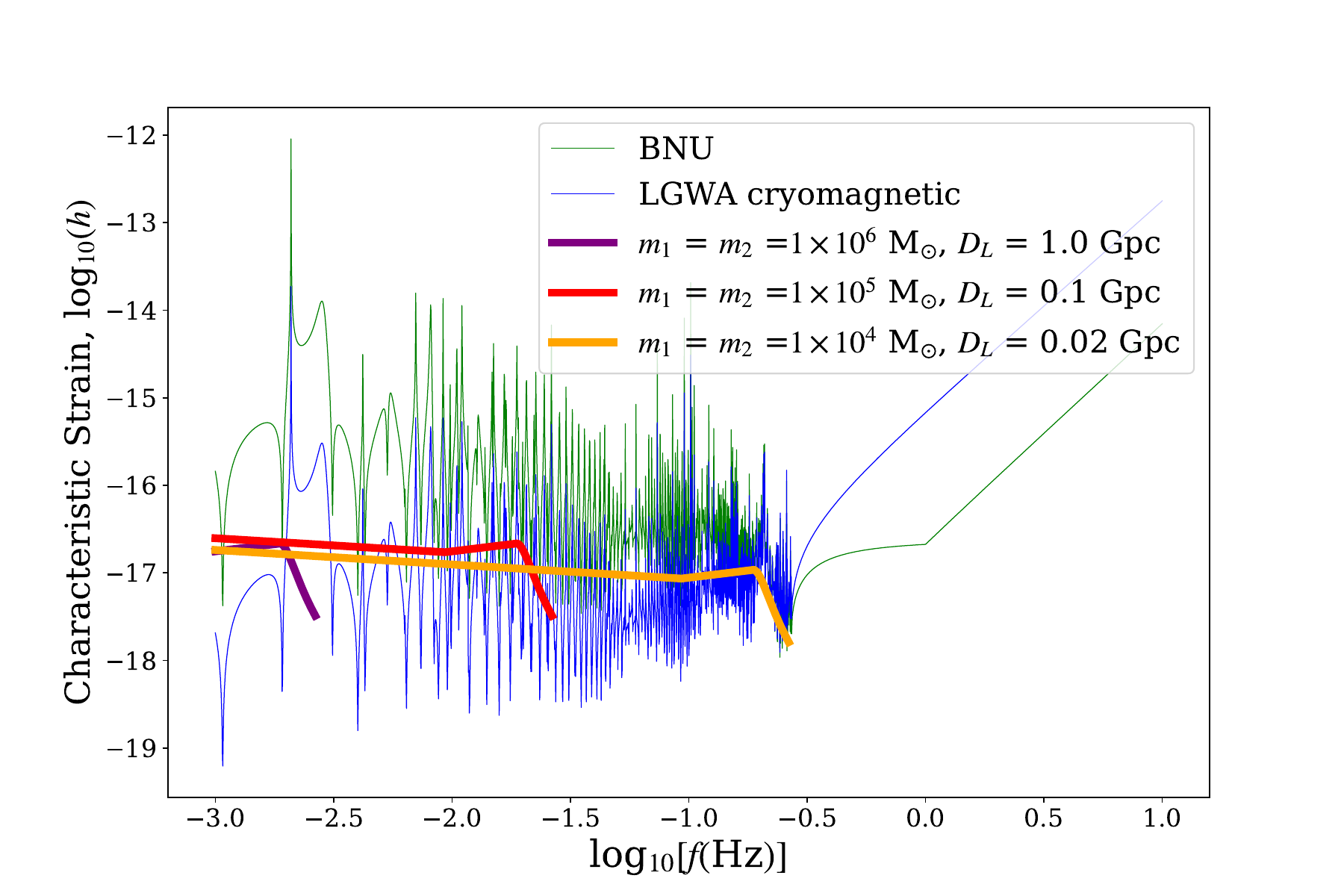}
\caption{\label{fig:4}GW characteristic strains for different instruments 
(thin lines) as well as representative astrophysical binary sources (thick lines). 
}
\end{figure}

Given the sensitivity of a lunar seismometer, $n_{f}$ (in unit of
$\mathrm{m\cdot s^{-2}/\sqrt{Hz}}$), we can now use our response functions
derived in coordinate $A$ to estimate the minimal detectable characteristic
strain of GW,
\begin{eqnarray}
    h_{n,r/h} = \frac{n_{f} }{\left ( 2\pi f \right )^{3/2} T_{r/h}^{A}  }, \label{eq:hn}
\end{eqnarray}
where the subscript $r/h$ means radial or horizontal directions. For example,
we consider two recently proposed lunar GW detectors, one based on the
cryomagnetic design from the LGWA project \cite{2021ApJ...910....1H} (hereafter
``LGWA cryomagnetic'') and the other operating at the moon surface temperature
proposed by Beijing Normal University \cite{2023SCPMA..6609513L} (hereafter
``BNU'').  The resulting sensitivities of the instruments to GWs (i.e.,
$h_{n,r/h}$) are shown in Figure~\ref{fig:4}. Most importantly,
because of the updated response curves, our sensitivity curves are flat at
$f<0.1$ Hz, while those in the previous works show a V-shape centered at
$0.1-1$ Hz.

To understand the effect of the new response curves on GW observation, we also
plot in Figure~\ref{fig:4} the characteristic strains of several IMBHBs and SMBHBs at
different distances. For simplicity, the two black holes in the binary are
considered to be equal, and the total mass is  $m = 2 \times 10^{6}$, $2 \times
10^{5}$, or $2 \times 10^{4} \text{M}_{\odot}$. We consider only circular
orbits, so the characteristic strain can be calculated with
\begin{eqnarray} 
h_{c}\left ( f \right )  = 2f\tilde{h} \left ( f \right )  ~,
\end{eqnarray}
where $\tilde{h} \left ( f \right ) $ is the Fourier transformation of the
public PhenomA template \cite{2007CQGra..24S.689A}. The luminosity distance
$D_{L}$ is chosen to be $1$, $0.1$, and $0.02$ Gpc, respectively, for the three
total masses given above. It is a bit arbitrary, chosen for demonstration only. In any case, the
characteristic strain is inversely proportional to $D_L$.

The signal-to-noise ratio (SNR) of a GW source is calculated according to the
standard definition, 
\begin{eqnarray}
    SNR^{2} = \int \mathrm{d} \left ( \mathrm{ln}f   \right ) \frac{h_{c}^{2} }{h_{n}^{2}}  ~.
\end{eqnarray}
Given the IMBHBs and SMBHBs specified above, the SNRs are $28.6, 39.7$, and
$29.8$ for LGWA cryomagnetic, and $0.42, 0.86$, and $1.02$ for BNU.  Although the SNRs 
for LGWA cryomagnetic are relatively high, the corresponding luminosity distances
are significantly lower than the previous estimations \cite{2021ApJ...910....1H,2023SCPMA..6609513L}
because here the response functions are updated using our own calculations.
The lower $D_{L}$ stresses the importance of improving the design of lunar seismometers
to further suppress the instrument noise.

\section{\label{sec:con}Summary and conclusions}

In this paper, we have revisited the theory of calculating the lunar response to GWs.
We clarified an ambiguity which exists in the literature about two response functions
derived from two viewpoints, one based on the Dyson
force (Equation~\ref{eq:elasGW1}) and the other from the ordinary tidal force
(Equation~\ref{eq:elasGW2}). We showed that 
the apparent difference between the two functions 
are caused by the choice of different coordinates.

Based on this understanding, we derived a concise and clear relationship
between the two functions (see Equation~\ref{eq:TFrela}). We verified this
analytical relationship by comparing the numerical response functions
calculated using, respectively, the Dyson and tidal forces   (see
Figure~\ref{fig:3}). A good agreement was found at lower
frequencies. The deviation at higher frequencies can be attributed to (i) a
truncation of the normal modes above a certain high value of $n$ in our
calculation and (ii) a limitation on the number of layers that we
implemented in the lunar model. 

The new response functions have a big impact on the detectability of GW sources
by future lunar seismometers. As Figure~\ref{fig:4} has shown, the sensitivity
to GWs given the current design of detectors flattens out between $10^{-3}$ and
$0.1$ Hz, making the detection of deci-Hertz GWs more challenging than
previously thought. In particular, to detect IMBHBs and SMBHBs, which are
important sources in the deci-Hertz GW band, it is essential to achieve in the
$10^{-3}-0.1$ Hz frequency band a sensitivity better than that of the
cryomagntic detector design by the LGWA project.  We believe our results will
help shaping up the scientific objectives of lunar GW observation, as well as
provide important constraints on the design of lunar GW detectors. 

Finally, we would like to point out that our response functions are
derived based on the current normal-mode formulation of the dynamical equation
of an elastic system in a GW field.  There are important aspects of the lunar
seismic response that are not captured by the current normal-mode model
according to the data from the Apollo seismic
observations~\cite{2021ApJ...910....1H,2022JGRE..12707222Z}. Further studies on the lunar structure
and lunar seismic response are urgently needed.

\begin{acknowledgments}
This work is supported by the National Key Research and Development Program of
China Grant No.s 2021YFC2203002 and 2023YFC2205802, the Beijing Natural Science Foundation No. 1242018, the Fundamental Research Funds for the Central Universities No. 310432103, and the National Natural Science Foundation of
China (NSFC) grants No. 11991053, 12073005, 12021003, and 42325406.
The authors would like to thank Yanbin Wang, Zijian Wang and Junlang Li for many discussions. We thank Biao Yang for detailed information about the LVZ structure, and also thank
Wuchuan Xu for his help in using the MINEOS package. Special thanks are due to our referee
for many constructive suggestions.
\end{acknowledgments}

\appendix

\section{\label{app:GWproj}Detailed derivations of Eq.~(\ref{eq:GWproj})} 

Using the definitions of three base vectors:
\begin{eqnarray}
    \hat{e}_{r} &&= \left ( \sin \theta \cos \varphi ,\sin \theta \sin \varphi, \cos \theta  \right ) \nonumber \\ \hat{e}_{\theta } &&= \left ( \cos \theta \cos \varphi ,\cos \theta \sin \varphi, -\sin \theta  \right )  \nonumber\\ \hat{e}_{\varphi  } &&= \left ( - \sin \varphi , \cos \varphi, 0  \right ) ~,
\end{eqnarray}
we have:

\begin{eqnarray}
    \hat{e}_{r}\cdot \epsilon \cdot \hat{e}_{r}&& = \sin^{2} \theta \left ( \sin 2\varphi + \cos 2\varphi\right ) \nonumber \\
    \hat{e}_{\theta }\cdot \epsilon \cdot \hat{e}_{r}&& = \sin \theta \cos \theta \left ( \sin 2\varphi + \cos 2\varphi\right ) \nonumber \\
    \hat{e}_{\varphi  }\cdot \epsilon \cdot \hat{e}_{r} &&= \sin \theta  \left (- \sin 2\varphi + \cos 2\varphi\right ) ~.
\end{eqnarray}

According to Ma19, the definition of the real spherical harmonics leads to the following results:

\begin{eqnarray}
    \mathcal{Y}_{2,2} \left ( \theta ,\varphi  \right ) &&= \frac{1}{4} \sqrt{\frac{15}{\pi } } \sin^{2}\theta \sin 2\varphi  \nonumber \\
    \mathcal{Y}_{2,-2} \left ( \theta ,\varphi  \right ) &&= \frac{1}{4} \sqrt{\frac{15}{\pi } } \sin^{2}\theta \cos 2\varphi  ~.
\end{eqnarray}

Considering the definition of $f^{m}$, we have:

\begin{eqnarray}
    \sum_{m} \mathcal{Y}_{2m} \left ( \theta ,\varphi  \right )f^{m} &&= \sin^{2}\theta\left ( \sin 2\varphi + \cos 2\varphi  \right ) \nonumber \\
    \sum_{m} \partial_{\theta }  \mathcal{Y}_{2m} \left ( \theta ,\varphi  \right )f^{m} &&= \sin2\theta\left ( \sin 2\varphi + \cos 2\varphi  \right ) \nonumber \\
    \sum_{m} \frac{\partial _{\varphi  }  \mathcal{Y}_{2m} \left ( \theta ,\varphi  \right )}{\sin \theta }f^{m} &&= 2\sin\theta\left ( -\sin 2\varphi + \cos 2\varphi  \right )  ~
\end{eqnarray}

Thus Eq.~(\ref{eq:GWproj}) has been proven by combining the above results.


\bibliography{main}

\end{document}